\documentclass[12pt,epsf,showkeys]{article}

\usepackage{graphicx,float}
\usepackage{mathrsfs,array,multirow}
\usepackage{amstext}
\usepackage{subfigure}
\usepackage{bm,latexsym,amsmath,amsfonts,amssymb}
\usepackage[usenames,dvipsnames]{color}
\usepackage{color}
\usepackage[colorlinks=true,linkcolor=blue]{hyperref}
\usepackage{soul}
\usepackage{epsfig}
\usepackage{subfigure}	

\newcommand{\be}{\begin{equation}}

\newcommand{\ee}{\end{equation}}

\newcommand{\ba}{\begin{array}}

\newcommand{\ea}{\end{array}}

\begin{document}
\begin{titlepage}
\vspace{.5in}
\begin{flushright}
CQUeST-2020-0655
\end{flushright}
\vspace{0.5cm}

\begin{center}
{\Large\bf Shadow cast by a rotating black hole with anisotropic matter}\\
\vspace{.4in}

  {$\mbox{Bum-Hoon Lee}^{\S\dag}$}\footnote{\it email: bhl@sogang.ac.kr},\,\,
  {$\mbox{Wonwoo Lee}^{\S}$}\footnote{\it email: warrior@sogang.ac.kr},\, \,
  {$\mbox{Yun Soo Myung}^{\P}$}\footnote{\it email: ysmyung@inje.ac.kr}\\

\vspace{.3in}

{\small \S \it Center for Quantum Spacetime, Sogang University, Seoul 04107, Korea}\\
{\small \dag \it Department of Physics, Sogang University, Seoul 04107, Korea}\\
{\small \P \it Institute of Basic Sciences and Department of Computer Simulation, Inje University,\\
Gimhae 50834, Korea}\\

\vspace{.5in}
\end{center}
\begin{center}
{\large\bf Abstract}
\end{center}
\begin{center}
\begin{minipage}{4.75in}

{\small \,\,\,\,
We obtain the shadow cast induced by the rotating black hole with an
anisotropic matter. A Killing tensor representing the hidden symmetry
is derived explicitly. The existence of a separability structure implies
complete integrability of the geodesic motion.  We analyze an
effective potential around the unstable circular photon orbits to show that  one side of the black hole is
brighter than the other side. Further, it is shown  that the inclusion of the anisotropic matter
($Kr^{2(1-w)}$) has an effect on  the observables of the shadow cast.  The shadow observables include
approximate shadow radius $R_s$, distortion parameter $\delta_s$,
area of the shadow $A_s$, and oblateness $D_{os}$.
}
\end{minipage}
\end{center}
\end{titlepage}

\newpage

\section{ Introduction \label{sec1}}

It is interesting to note that a black hole is one of the fascinating and mystical objects in the universe.
Its physical meaning and existence as a real object  have developed over the past
century~\cite{Schwarzschild:1916uq, Oppenheimer:1939ue, Kerr:1963ud, Newman:1965my, Penrose:1964wq,
Penrose:1969pc, Ruffini:1971bza, Broderick:2015tda}.
It is fair to say that black holes do not seem to be directly observed.
Therefore, they have been proved by
indirect observations like the deflection of light rays due to the spacetime
curvature~\cite{Einstein:1956zz, Reber1944ap, Schmidt:1963wkp, Narayan:2013gca},
or the gravitational wave by coming from the mergers of compact binary sources \cite{Abbott:2016nmj,Abbott:2017vtc}.

Interestingly, the black hole has recently gained the most attention thanks to the observational
reports on the shadow cast induced by the supermassive black hole~\cite{Kormendy:1995er,
Ghez:2008ms, Gillessen:2008qv} at the center
of the M87 galaxy obtained by the Event Horizon Telescope
collaboration~\cite{Akiyama:2019cqa, Akiyama:2019bqs, Akiyama:2019fyp}.
The shadow image indicates the bright ring surrounding a dark region in the celestial sphere.
The dark area describing the black hole shadow has the boundary between capture orbits and
scattering orbits by photons  in which photons are coming from both the accreting disc and the
light source located behind a black hole, and they reach a distant observer~\cite{Bardeen:1973tla,
CunBar1973, Young:1976zz, Luminet:1979nyg, Falcke:1999pj, Cunha:2018acu, Dokuchaev:2018kzk, Narayan:2019imo}.

Let us propose an astrophysical black hole residing in the background of matters. In this
case, it is appropriate to consider a realistic black hole surrounded by a matter field or dark
matter. It is well-known that dark matter~\cite{Zwicky:1933gu, Rubin:1970zza, Aghanim:2018eyx, Kang:2019uuj}
makes up about $27\%$ of the universe and more than $90\%$ of the matter in  the Milky Way.
To model a black hole in the galaxy, it is reasonable to consider the black hole coexisting
with matter field or dark matter~\cite{Cho:2017nhx, Cho:2017vhl, Kim:2019hfp}.

Analyzing the geodesic motion of a photon  outside the black hole horizon is an important
matter when studying  astrophysical objects exposed to observations. Theoretically, studying
the geodesic motion of a photon may provide  a clear picture of geometrical properties for the
neighborhood of a rotating black hole~\cite{Dymnikova1986}. One may construct astrophysical models exposed
to observations by describing the geodesic motion~\cite{Bardeen:1973tla}. Actually, the black hole shadow
could be investigated by analyzing the null geodesics around the rotating black hole.

The black hole shadow could be used to measure physical parameters (mass, charge, angular
momentum, inclination angle, structure of the accretion disk). All parameters may include the
effects of the black hole surrounded with matter fields or those in modified gravity theories. For
this reason, the shadow cast has been extensively investigated in gravity theory with/without
matter fields~\cite{Bardeen:1973tla, Young:1976zz, Vries2000,
Hioki:2009na, Konoplya:2019sns, Atamurotov:2015nra, Abdujabbarov:2015xqa, Abdujabbarov:2016hnw,
Hou:2018bar, Stuchlik:2018qyz, Haroon:2018ryd, Jusufi:2019nrn, Wei:2019pjf, Roy:2019esk, Vagnozzi:2020quf, Chang:2020miq,
Badia:2020pnh, Dokuchaev:2020wqk, He:2020dfo, Zhang:2020xub, Chowdhuri:2020ipb, Bambhaniya:2021ybs},
other rotating objects~\cite{Nedkova:2013msa, Bambi:2013nla, Amir:2018pcu, Bambi:2019tjh},
and modified gravity theories~\cite{Hioki:2008zw, Amarilla:2011fx, Wei:2013kza, Cunha:2015yba,
Younsi:2016azx, Sharif:2016znp, Shaikh:2019fpu, Vagnozzi:2019apd, Allahyari:2019jqz, Kumar:2020pol,
EslamPanah:2020hoj, Khodadi:2020jij, Long:2020wqj, Contreras:2020kgy}.

Recently, two of us have obtained the rotating black hole solution with an anisotropic
matter~\cite{Kim:2019hfp} where the  anisotropic matter with parameters $w$ and $K$ may describe both an extra
U(1) field~\cite{Schee:2008kz, Amarilla:2011fx, Astefanesei:2019pfq, Zou:2019ays, Myung:2020dqt}
and diverse dark matters. It is shown that this black hole geometry
can affect the shadow  when comparing  with  Kerr and Kerr-Newman black holes~\cite{Badia:2020pnh}.
We are also interested in studying the shadow cast  induced by the black hole with an anisotropic
matter as a subsequent study.
For this purpose, we will derive the Killing tensor and  find the separability structure to guarantee the
integrability of the geodesic motions. Also, we investigate an effective potential for unstable circular photon orbits to show that
one side of the black hole is brighter than the other side.
As additional  shadow observables, we analyze the approximate shadow radius and distortion parameter.
We hope  that all may have a complementary relationship with one another, to describe the nature.

In this work, we wish to  focus on studying the shadow cast induced by the black hole with
an anisotropic matter.  The paper is organized as follows. In Sec.\ $2$, we  explore the
symmetry of the rotating black hole geometry~\cite{Frolov:2017kze}. A Killing tensor representing the hidden
symmetry is constructed and the separability structure exists.
In Sec.\ $3$, we derive the geodesic equation by adopting the Hamilton-Jacobi formalism.
To get the information on the boundary of the shadow cast, we study the radial null geodesic motion by making use of the effective potential.
In Sec.\ $4$, we employ a backward ray-tracing algorithm~\cite{Vincent:2011wz, James:2015yla, Cunha:2016bpi} to analyze  the shadow cast induced by a rotating black hole with an anisotropic matter field described by two parameters $w$ and $K$.
We present the apparent shape of shadow and  observables characterizing the shadow. We observe that the anisotropic
matter field with $w$ influences on the observables.   Finally, we summarize our results and discuss on relevant matters in Sec.\ $5$.

\section{Symmetry and separability structure \label{sec2}}
First of all, we consider the action
\begin{equation}
I=\int d^4x \sqrt{-g}\Big[\frac{1}{16\pi}(R-F_{\mu\nu}F^{\mu\nu})+{\cal L}_{\rm am}\Big] + I_{b},
\end{equation}
where ${\cal L}_{\rm am}$ describes effective anisotropic matter fields and $I_{b}$ is the boundary term~\cite{Gibbons:1976ue, Hawking:1995ap}.
The rotating black hole solution with an anisotropic matter is obtained by applying the Newman-Janis algorithm to a static spherically symmetric solution as~\cite{Kim:2019hfp}
\begin{equation}
ds^2=  - F(r, \theta) dt^2 -2[1-F(r, \theta)]a\sin^2\theta dt d\phi  +
\frac{\Sigma}{\rho^2} \sin^2\theta d\phi^2 + \frac{\rho^2}{\triangle} dr^2 + \rho^2 d\theta^2 \,,
\label{metric}
\end{equation}
where
\begin{eqnarray}
 F(r, \theta)&=&1-\frac{2Mr-Q^2+Kr^{2(1-w)}}{\rho^2} ,\quad a=\frac{J}{M},\quad\rho^2=r^2+ a^2\cos^2\theta, \nonumber \\
 \triangle & =& r^2 + a^2 + Q^2 -2Mr - K r^{2(1-w)}, \nonumber \\
 \Sigma&=& \rho^2(r^2+a^2) + (2Mr -Q^2 + K r^{2(1-w)})a^2\sin^2\theta. \label{F-rho}
\end{eqnarray}
It is important to point out that the parameters $K$ and $w$  control the density
and anisotropy of the fluid matter surrounding the black hole.
The  $K=0$ case leads to the Kerr-Newman black hole and $K=0$ with $Q=0$ corresponds to the Kerr black hole, regarding as two reference black holes.  This  includes the rotating version of the Reissner-Nordstr\"om black hole with the constant scalar hair~\cite{Zou:2019ays}
characterized by $K <0 $, $w=1$, and  $a=0$.  The metric is asymptotically flat for $w>0$ only.  For  $0 \leq w \leq 1/2$, the energy
density is not localized sufficiently such  that the total   energy diverges~\cite{Kim:2019hfp}.
Thus, we  consider  the case with $w > 1/2$ to obtain  the geometry including a finite total energy with  asymptotically flat spacetime.
We allow $K$ to take either sign for  representing diverse matters surrounding the rotating black hole.

At this stage, we mention that the event (outer)  horizon ($r_H$) for the spacetime (\ref{metric}) is located at the largest radius as a solution to $\triangle=0$, leading to the event horizon for Kerr-Newman black hole with  $K=0$.

We are interested particularly in  the ergosphere for later use in Sec.\ \ref{sec3}.
The ergosphere is defined as a region  between  surface of static limit (infinite redshift)  and outer horizon~\cite{Ruffini:1970sp}.
On the surface, the timelike Killing vector becomes null-like as $g_{tt}=0$.
Let us consider  a photon emitted in the $\phi$ direction at radius $r$ without $r$ and $\theta$  momentum components.
 From the condition of the null  trajectory, one obtains  two angular velocities
\begin{equation}
\omega_{\pm}= \frac{-g_{t\phi}\pm\sqrt{\triangle\sin^2\theta}}{g_{\phi\phi}} \,,
\label{angvel}
\end{equation}
where $\triangle\sin^2\theta=g^2_{t\phi}-g_{tt}g_{\phi\phi}$.
Outside the event horizon ($\triangle >0$), two roots
$\omega_{\pm}$ are alive, while  there are no real roots inside the horizon. At the static limit surface satisfying  $g_{tt}=0$ and $g_{t\phi} < 0$,
one finds $\omega_+ =-\frac{2g_{t\phi}}{g_{\phi\phi}}$ and $\omega_- =0$.
Inside the ergosphere ($g_{tt}>0$), one has  $\omega_- >0$. Therefore, all particles are necessarily corotating
with the rotating black hole. On the event horizon, the two angular velocities appear the same as
$\omega_{\pm}=\Omega|_{r=r_H}= \frac{a}{r^2_H + a^2}$.

Now, we examine the symmetry and separability structure of the rotating black hole geometry.
There are  explicit and hidden symmetries  related to  Killing vectors
and  Killing tensors, respectively~\cite{Eisenhart1949, Frolov:2017kze}.
Killing vectors correspond to the generators of isometries for spacetime geometry.
The  geometry of  (\ref{metric}) implying a stationary axisymmetry,  admits two commuting Killing vectors
$\xi^{\mu}_{(t)}=\delta^{\mu}_t$ and $\xi^{\mu}_{(\phi)}=\delta^{\mu}_{\phi}$~\cite{Carter:1971zc}.
On the other hand, Killing tensors correspond to a symmetric generalization of Killing vectors.
A hidden symmetry represents the geometric structure of spacetime encoded
in Killing tensors. This implies that an additional integral of the  motion is quadratic in momentum as
${\cal K}^{\mu\nu}p_{\mu}p_{\nu}$~\cite{Walker:1970un}.
It is known that one of the Killing tensors is the metric tensor, having the structure of  $g^{\mu\nu}p_{\mu}p_{\nu}=-m^2$.

We consider the null tetrad~\cite{Kinnersley:1969zza, Kim:2000gy} to construct  the Killing tensor.
The null tetrad consists of two real null vectors  $l^{\mu}$ and $n^{\mu}$ and two complex null vectors $m^{\mu}$ and ${\bar m}^{\mu}$.
They satisfy $l^{\mu}n_{\mu}=-1$, $m^{\mu}{\bar m}_{\mu}=1$, and
$l^{\mu}m_{\mu}=l^{\mu}{\bar m}_{\mu}=n^{\mu}m_{\mu}=n^{\mu}{\bar m}_{\mu}=0$.
For our purpose, we introduce   an inverse metric for (\ref{metric})
\begin{displaymath}
g^{\mu\nu}= \left(\begin{array}{cccc}
-\frac{\Sigma}{\triangle\rho^2} & 0 & 0  &  -\frac{a(1-F)}{\triangle} \\
0 & \frac{\triangle}{\rho^2} & 0 & 0 \\
0  & 0 & \frac{1}{\rho^2} & 0 \\
-\frac{a(1-F)}{\triangle} & 0 & 0 & \frac{\triangle-a^2\sin^2\theta}{\triangle\rho^2\sin^2\theta}
\end{array}  \right).
\label{metinve}
\end{displaymath}
Then, $ g^{\mu\nu}$ can be expressed in terms of the null tetrad as
\begin{equation}
g^{\mu\nu}= -l^{\mu}n^{\nu} -n^{\mu}l^{\nu} + m^{\mu}\bar{m}^{\nu} +\bar{m}^{\mu}m^{\nu} \,, \label{metricstandecom}
\end{equation}
where the null vectors are given by
\begin{eqnarray}
&&l^{\mu} =\frac{1}{\triangle}\left[(r^2+a^2) \delta^{\mu}_0 + \triangle \delta^{\mu}_1 + a \delta^{\mu}_3  \right] \,,~ n^{\mu}=\frac{1}{2\rho^2}\left[(r^2+a^2)\delta^{\mu}_0 - \triangle \delta^{\mu}_1 + a \delta^{\mu}_3  \right] \,, \nonumber \\
&& m^{\mu} =\frac{1}{\sqrt{2} (r+ia\cos\theta)}\left[ia \sin\theta \delta^{\mu}_0 + \delta^{\mu}_2 + \frac{i}{\sin\theta} \delta^{\mu}_3  \right] \,, \nonumber \\
&& {\bar m}^{\mu} =\frac{1}{\sqrt{2} (r-ia\cos\theta)}\left[-ia \sin\theta \delta^{\mu}_0 + \delta^{\mu}_2 - \frac{i}{\sin\theta} \delta^{\mu}_3  \right] \,.
\end{eqnarray}
Importantly, a quadratic Killing tensor is constructed  by making use of these null vectors~\cite{Walker:1970un, Chandrasekhar:1985kt}
\begin{eqnarray}
{\cal K}^{\mu\nu} &=&2\rho^2 m^{(\mu}{\bar m}^{\nu)} -a^2\cos^2\theta g^{\mu\nu}  \nonumber \\
 &=& a^2 \sin^2 \theta \delta^{\mu}_{0}\delta^{\nu}_{0} + 2a\delta^{(\mu}_{0}\delta^{\nu)}_{3} +\frac{\delta^{\mu}_{3}\delta^{\nu}_{3}}{\sin^2\theta} + \delta^{\mu}_{2}\delta^{\nu}_{2} -a^2 \cos^2\theta g^{\mu\nu} \,,
\label{Killingten}
\end{eqnarray}
which  satisfies $\nabla^{(\alpha}{\cal K}^{\mu\nu)}=0$ and ${\cal K}^{\mu\nu}={\cal K}^{(\mu\nu)}$.

Let us consider the separability structure~\cite{Benenti1979, Demianski1980} which describes the separation
of variables in the Hamilton-Jacobi formalism.
We are aware that there exist two Killing vectors ($\xi^{\mu}_{(t)}$ and $\xi^{\mu}_{(\phi)}$) satisfying $\nabla^{(\alpha}\xi^{\mu)}=0$,
and two Killing tensors (${\cal K}^{\mu\nu}$ and $g^{\mu\nu}$).
The metric tensor also satisfies $\nabla^{(\alpha}g^{\mu\nu)}=0$ and $g^{\mu\nu}=g^{(\mu\nu)}$.
One can show that two Killing tensors mutually commute under the Schouten-Nijenhuis bracket
\begin{equation}
[{\cal K}^{\mu\nu}, g^{\mu\nu}]_{\rm SN}=2{\cal K}^{\alpha(\mu} \nabla_{\alpha} g^{\nu\beta)}- 2g^{\alpha(\mu} \nabla_{\alpha}
{\cal K}^{\nu\beta)}=0 \,,
\end{equation}
which is regarded as   an alternative form of the Killing tensor equation.
Also, the  Killing tensors and vectors satisfy
\begin{equation}
[\xi^{\mu}_{(t)}, {\cal K}^{\mu\nu}]_{\rm SN}={\cal L}_{\xi^{\mu}_{(t)}} {\cal K}^{\mu\nu}=
[\xi^{\mu}_{(\phi)}, {\cal K}^{\mu\nu}]_{\rm SN}=[\xi^{\mu}_{(t)}, g^{\mu\nu}]_{\rm SN}=[\xi^{\mu}_{(\phi)}, g^{\mu\nu}]_{\rm SN}=[\xi^{\mu}_{(t)}, \xi^{\mu}_{(\phi)}]_{\rm SN}=0 \,.
\end{equation}
Therefore, we prove that  the rotating black hole geometry admits the separability structure.
Its existence  guarantees a complete integrability of the geodesic motions.

\section{Geodesic motions \label{sec3}}

In this section, we investigate the geodesic motions~\cite{Gwak:2008sg, Gwak:2011qs, Lee:2017fbq, Rizwan:2020tpc}.
For the static spherically symmetric black hole,
one can construct four  integrals of geodesic motion: test particle's energy,
projection of angular momentum to an arbitrary axis,  square of
total angular momentum, and  normalization of the four-velocity.  These are conserved along the
geodesics. Therefore, the geodesic equation becomes completely separable.
For an  axisymmetric rotating black hole, the total angular momentum is not conserved.
However, determining the orbit of a test particle  is a necessary step  to find
four integrals of the geodesic motion.
In this direction, Carter has obtained the fourth constant by performing  the separation of
variables in the Hamilton-Jacobi formalism~\cite{Carter:1968rr}.
Here, we wish to  construct four independent integrals of the  geodesic motion
by making use of  two Killing vectors and two Killing tensors.

The four integrals of the motion are given by  two conserved quantities related to Killing vectors
\begin{eqnarray}
\xi^{\mu}_{(t)}p_{\mu} &=& -E= - F(r, \theta) \dot{t} - [1-F(r, \theta)]a\sin^2\theta  \dot{\phi} \,, \nonumber \\
\xi^{\mu}_{(\phi)}p_{\mu} &=&L_z= - [1-F(r, \theta)]a\sin^2\theta \dot{t} + \frac{\Sigma \sin^2\theta}{\rho^2} \dot{\phi} \,,
\label{killangm}
\end{eqnarray}
and  two conserved quantities related to  Killing tensors
\begin{eqnarray}
g^{\mu\nu} p_{\mu}p_{\nu} &=& -m^2 \,, \nonumber \\
{\cal K}&\equiv& {\cal K}^{\mu\nu} p_{\mu}p_{\nu} = p^2_{\theta} + (L_z - a E\sin^2\theta)^2/\sin^2\theta + a^2 m^2 \cos^2\theta \,. \label{constantK}
\end{eqnarray}
Here, we investigate the geodesic motion around the rotating black hole  by following the procedure of
the separation of variables in the Hamilton-Jacobi formalism~\cite{Carter:1968rr}.
In this case, Carter's fourth constant of the motion is given by
\begin{eqnarray}
{\cal Q} =p^2_{\theta} + \cos^2\theta [a^2 (m^2 - E^2) + L^2_z/\sin^2\theta] \,,
\end{eqnarray}
which represents  the separation constant  for the $r$ and $\theta$-directions  of null geodesics.
The constant of the motion is given by
\begin{eqnarray}
{\cal Q} + (L_z - aE)^2 =p^2_{\theta} + (L_z - a E\sin^2\theta)^2/\sin^2\theta + a^2 m^2 \cos^2\theta  \,,
\label{caterKilling}
\end{eqnarray}
indicating that  ${\cal Q}$  may be  negative but ${\cal K}$ is always non-negative. Equation (\ref{caterKilling})
has the same form as Eq.\ (\ref{constantK}) implying  ${\cal K} \equiv {\cal Q} + (L_z - aE)^2$.

We notice that the geodesic motion of a neutral particle is described by the Hamilton-Jacobi equation.
The geodesic equations as four first-order differential equations are found to be
\begin{eqnarray}
&& \rho^2 p^t \equiv \rho^2 \frac{dt}{d\lambda}  = -a(aE\sin^2\theta-L_z) + \frac{(r^2 +a^2)P(r)}{\triangle}  \,, \label{HJeq-t} \\
&& \rho^2 p^{\phi} \equiv \rho^2 \frac{d\phi}{d\lambda} =  - \left(a E - \frac{L_z}{\sin^2\theta} \right) + \frac{a P(r)}{\triangle} \,, \label{HJeq-phi} \\
&& \rho^2 p^r \equiv \rho^2 \frac{dr}{d\lambda} = \pm \sqrt{R(r)} = \triangle p_r\,, \label{HJeq-r}  \\
&& \rho^2 p^{\theta} \equiv \rho^2 \frac{d\theta}{d\lambda} = \pm \sqrt{\Theta(\theta)} \,.  \label{HJeq-the}
\end{eqnarray}
Here $+(-)$ in Eqs.\ (\ref{HJeq-r}) and (\ref{HJeq-the}) correspond to the outgoing
(ingoing) geodesics and
\begin{eqnarray}
&&\Theta(\theta) = {\cal Q} - \cos^2\theta [a^2 (m^2 - E^2) + L^2_z/\sin^2\theta], \label{the-r-p-func1} \\
&&R(r) = P^2(r) - \triangle [m^2 r^2 +(L_z-a E)^2 + {\cal Q} ]\,,\label{the-r-p-func2} \\
&&P(r)= (r^2+a^2)E - aL_z \,. \label{the-r-p-func3}
\end{eqnarray}
The geodesic motion of a test particle is not confined in a plane, implying that two
effective potentials for the radial and latitudinal motions should be examined separately.

Now we are in a position to study the null geodesic motion with $m=0$.
It is noted that  the angular size of a light source is much bigger than the angular size of the black hole.
They are two important null geodesic motions named as  principal congruences and unstable circular orbits.
We focus on unstable circular orbits  and then, mention  the principal congruences briefly.
The radial equation for null geodesic is given by
\begin{equation}
\frac{1}{2} \left(\frac{dr}{d\lambda} \right)^2  + V_{\rm eff} (r)=0 \,,
\end{equation}
where the effective potential in the equatorial plane is given by
\begin{eqnarray}
V_{\rm eff} (r)= - \frac{((r^2+a^2)^2-\triangle a^2)E^2 - 2a L_z(r^2+a^2-\triangle)E + (a^2-\triangle)L^2_z}{2r^4}\,.
\label{eff-pot-null}
\end{eqnarray}
Here, we note that  Carter's constant of the motion  $({\cal Q})$  disappears. The local maximum in the effective potentials determines the radii
of unstable circular photon orbits.  For $K=0$, this reduces to  the Kerr-Newman potential
\begin{eqnarray}
V^{\rm KN}_{\rm eff}(r)= - \frac{[r^2(r^2+a^2)+ a^2(2Mr -Q^2)]E^2-2aL_z(2Mr-Q^2) E +(-r^2+2Mr-Q^2)L^2_z}{2r^4}.
\label{Keff-pot-null}
\end{eqnarray}

\begin{figure}
\begin{center}
\subfigure[Effective potentials]
{\includegraphics[width=3.2in]{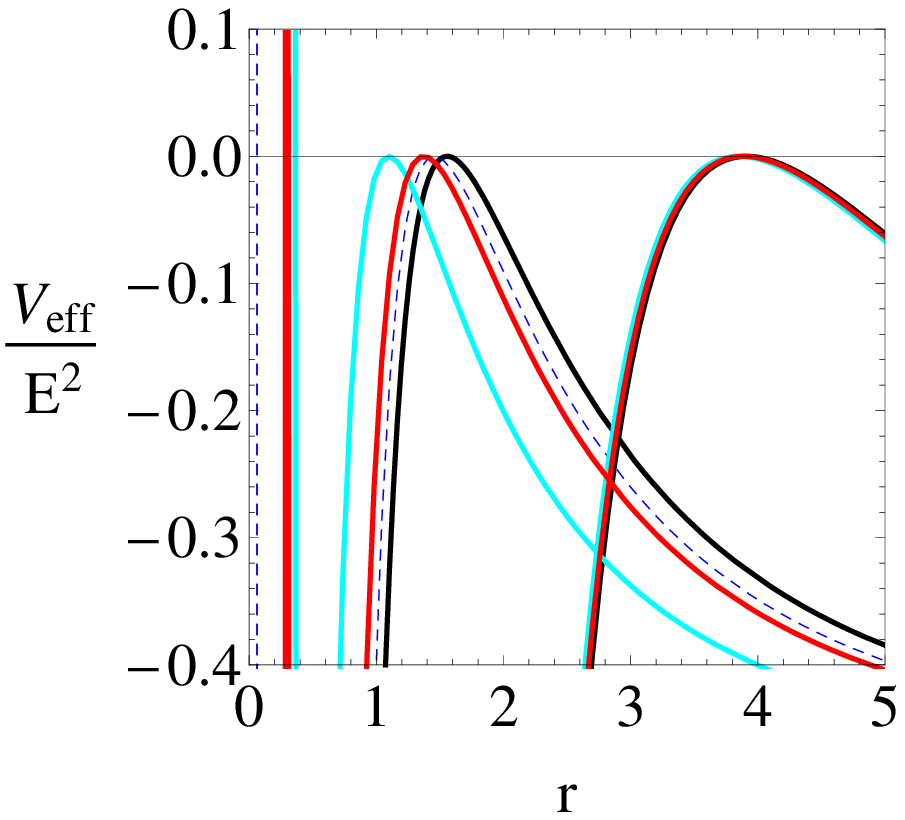}}
\subfigure[Locations of unstable circular orbits]
{\includegraphics[width=2.7in]{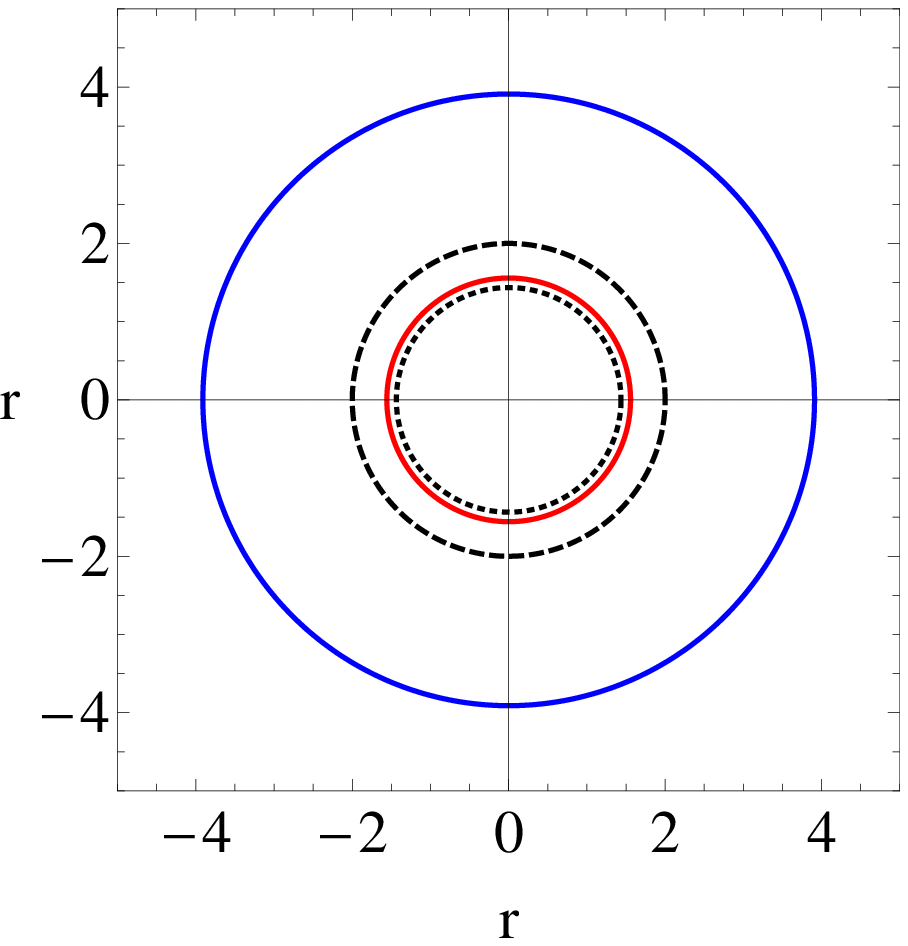}}
\end{center}
\caption{\footnotesize{(color online).
Effective potentials and locations of unstable circular orbits in the equatorial plane.
}}
\label{Effective_potential}
\end{figure}

For unstable circular orbits, we have to  take into account  $V_{eff}=0$ and $\frac{dV_{eff}}{dr}=0$. Solving these,  the location of a peak is determined as
\begin{eqnarray}
r_{uco}&=& \frac{L_z -aE}{L_z +aE}\Big[3M - \frac{2Q^2}{r_{uco}} + K(1+w)r^{(1-2w)}_{uco}\Big] \,.
\end{eqnarray}
For $K=0$, this reduces to  the Kerr-Newman case.
For photon circular orbits around the extremal Kerr black hole with $a=M$, one has either  $r_{\rm uco-}=M$ (corotating)
or $r_{\rm uco+}=4M$ (counterrotating case).

Carter's constant determines  the test particle's motion in the $\theta$-direction \cite{Wilkins:1972rs, Teo2003}.
A physically arrowed region is  defined by  $\Theta(\theta) \geq 0$.
The orbits cross the equatorial plane repeatedly for ${\cal Q} > 0$, while they remain in the equatorial plane
for ${\cal Q} = 0$. We have the condition of ${\cal Q} < 0$ for the principal null congruences.

We introduce the dimensionless quantities  ($\bar{a}=a/M$, $\bar{Q}=Q/M$, $\bar{K}=K/M^{2w}$, $\bar{r}=r/M$).
Hereafter, we will remove the bar for simplicity.

Figure \ref{Effective_potential}(a) shows the effective potential for both corotating and counterrotating cases
in the equatorial plane. The left concave curves correspond to the former case, while the right concave curves to
the latter case. As reference curves, black curves correspond to the Kerr one with $a=0.9$, blue dashed
curves to Kerr-Newman one \cite{Zajacek:2018ycb} with $Q=0.25$. For our work,
we note that red curves denote  the case with $a=0.9$, $Q=0$
$w=3/2$, $K=-0.13$ while  cyan curves  represent the case with  $a=0.9$, $Q=0.25$, $w=3/2$, $K=-0.13$.

Figure \ref{Effective_potential}(b) shows the locations of unstable circular orbits.
The black dotted (dashed) circle indicates the location of  outer  horizon (static limit).
The red circle denotes the unstable circle orbit for the corotating case, while the blue circle represents  the counterrotating case.
We note that the unstable circle orbit for the corotating case is located
within the ergosphere, implying that the angular velocity of null rays for the counterrotating case could
change the sign in the ergosphere and they move along unstable circle orbit.
Hence, the number of photons for corotating case is larger than those for counterrotating case
when reaching the distant observer.

Now, let us  consider null geodesics along $\theta=\theta_c=$const plane.
Considering that $\dot r$ and $\dot \theta$ must be real on  the geodesics in Eqs.\ (\ref{HJeq-r}) and (\ref{HJeq-the}),
one has  $R(r) \geq 0$ and $\Theta(\theta) \geq 0$ in Eq.\ (\ref{the-r-p-func1}). $\Theta(\theta)$  is rewritten as
\begin{eqnarray}
&&\Theta(\theta) = {\cal Q} + (a^2E^2 + L^2_z) - \left(a^2E^2\sin^2\theta + \frac{L^2_z}{\sin^2\theta}\right) \,.
\label{}
\end{eqnarray}
For $\theta=\theta_c$, one has
\begin{equation}
\Theta(\theta)|_{\theta=\theta_c} = 0\,,\quad \partial_{\theta}\Theta(\theta)|_{\theta=\theta_c}=0 \,.
\end{equation}
Equivalently, one finds
\begin{equation}
\frac{L_z}{E} = a\sin^2\theta_c\,,\quad \frac{{\cal Q}}{E^2}= -a^2 \cos^4\theta_c \,.
\label{prinnucon}
\end{equation}
If  Eq.\ (\ref{prinnucon}) is satisfied, the solution to the geodesic equations
provides  the principal null geodesic.
From Eqs.\ (\ref{HJeq-t}), (\ref{HJeq-phi}), and (\ref{HJeq-r}) with Eq.\ (\ref{prinnucon}), we get
\begin{equation} \label{coordi}
\frac{dt}{dr} = \pm \frac{r^2+a^2}{\triangle}\,,\quad \frac{d\phi}{dr} = \pm \frac{a}{\triangle} \,.
\end{equation}
The outgoing(+)/ingoing($-$) radial congruences correspond to the two principal null congruences
in Boyer-Lindquist coordinates. We may refer  Refs.\ \cite{Petrov:2000bs, Pirani:1956wr, Penrose:1960eq, Misner:1974qy}
for discussion on the Petrov-Pirani-Penrose classification.

For the Kerr-Newman case, (\ref{coordi})  can be  explicitly solved to give
\begin{equation}
\pm t = r+ \frac{r^2_+ + a^2}{r_+ - r_-}\ln|r-r_+| - \frac{r^2_- + a^2}{r_+ - r_-}\ln|r-r_-|+ {\rm const}  \,.
\end{equation}
and
\begin{equation}
\pm \phi= \frac{a}{r_+ - r_-} \ln{\big|\frac{r-r_+}{r-r_-}\big|} + {\rm const} \,.
\end{equation}
Finally, we mention  that the ingoing principal null congruence crosses the event horizon when using Kerr (Edding-Finkelstein) coordinates.

\section{Shadow cast \label{sec4}}

The black hole shadow corresponds to the gravitational capture cross-section of photons.
We adopt the backward ray-tracing algorithm to obtain a connection between impact parameters
and celestial coordinates~\cite{Psaltis:2010ww, Cunha:2016bpi}.
Finally, we wish to show the shadow cast induced  by the rotating black hole with an anisotropic matter field.

\subsection{Backward ray-tracing algorithm \label{sec4-1}}

The ray-tracing algorithm is designed  for generating an image by tracing the path of light scattered
off the surface of an object. There are two algorithms named forward and backward ones.
The former corresponds to the method in which light rays emitted from a source are scattered  off an object,
enter an optical device, and finally generate the image.  The latter denotes  the opposite
travel direction of light rays. We  may trace individual light rays backward in time from an image plane.

We set up the image plane being  perpendicular to the observer's line of sight to describe the black hole
shadow in the celestial sphere. It is assumed that the plane (or observer) is located at an infinitely large distance
from the light source with an inclination angle. The light source is considered as both the photons
passing near a black hole and emitting from an accretion disk   from the observer's view.

We should choose a proper set of tetrad basis vectors ($e^{\mu}_{\hat{a}}$) to obtain the locally measured
quantities by projecting  photon's momentum $p^{\mu}$.
This  is related  to choosing  the locally nonrotating frame
or the reference frame of zero angular momentum observer (ZAMO) \cite{Bardeen:1972fi, Frolov:1998wf}.
We note that the ZAMO frame is a local inertial  frame.  An observer at rest in the ZAMO frame acquires an angular
velocity as an effect of frame-dragging caused by  gravitational field of a rotating black hole.
A useful set of the tetrad is given by
\begin{eqnarray}
e^{\mu}_{\hat{t}}&=& \left(\frac{1}{\rho}\sqrt{\frac{\Sigma}{\triangle}}, 0, 0, \frac{(1-F)a\rho}{\sqrt{\Sigma\triangle}} \right)\,,~~~~e^{\mu}_{\hat{r}}= \frac{\sqrt{\triangle}}{\rho}(0,1,0,0) \,, \nonumber \\
e^{\mu}_{\hat{\theta}}&=&\frac{1}{\rho}(0,0,1,0)\,, \quad e^{\mu}_{\hat{\phi}} =\frac{\rho}{\sqrt{\Sigma} \sin\theta}(0,0,0,1) \,. \label{otetrad}
\end{eqnarray}
Then, the locally measured energy and the  momentum for photon are given by
\begin{eqnarray}
p^{\hat{t}}&=& \frac{1}{\rho}\sqrt{\frac{\Sigma}{\triangle}} E - \frac{(1-F)a\rho}{\sqrt{\Sigma\triangle}} L_z \,,
~~~~p^{\hat{r}}= \frac{\rho}{\sqrt{\triangle}} p^r \,, \nonumber \\
p^{\hat{\theta}}&=& \rho p^{\theta}\,, ~~~~ p^{\hat{\phi}}=  \frac{\rho}{\sqrt{\Sigma} \sin\theta} L_z \,,
\label{loenmon}
\end{eqnarray}
with $p_t=-E$ and $p_{\phi}=L_z$.
For a static observer at spatial infinity, the momentum of
photon  turns out to be ($w > 1/2$)
\begin{eqnarray}
p^{\hat{t}}&\rightarrow&  E\,, ~~~~ p^{\hat{r}} \rightarrow p^r \,, ~~~~ p^{\hat{\theta}}\rightarrow r p^{\theta}\,, ~~~~
p^{\hat{\phi}} \rightarrow \frac{L_z}{r \sin\theta}\,.
\label{asymenmon}
\end{eqnarray}

\subsection{Impact parameters \label{sec4-2}}

We wish to construct the Cartesian-like coordinates in the image plane to show  the apparent shape
of a black hole shadow composed  of individual photons. Actually, these are  the observation
angles of $\alpha$ and $\beta$~\cite{Psaltis:2010ww, Cunha:2016bpi}.
We expect that these  are regarded as  coordinate axes in the plane at spatial infinity.
In order to get at these, we introduce two impact parameters defined as
$\bar{\alpha} \equiv r_o \alpha$ and $\bar{\beta} \equiv -r_o \beta$,
in which $r_o$ is computed at   the position of the observer. These impact parameters correspond to
the coordinate axes when taking $r_o \rightarrow \infty$. Hereafter, we will
remove the bar for simplicity.

We note that $R(r)$ and $\Theta(\theta)$ must be non-negative.  For the photon motion, we  have
\begin{eqnarray}
\frac{R(r)}{E^2} &=& [(r^2+a^2)-a\xi]^2- \triangle[\eta + (\xi-a)^2] \geq 0\,,  \\
\frac{\Theta(\theta)}{E^2} &=& \eta + (\xi-a)^2 - \left(\frac{\xi}{\sin\theta} -a\sin\theta \right)^2 \geq 0 \,,
\end{eqnarray}
with $\xi=L_z/E$ and $\eta={\cal Q}/E^2$.

In  general rotating spacetime, the unstable circular photon orbits    satisfy $R|_{r=r_{uco}}=0$,
$R'|_{r=r_{uco}}=0$,  and $R''|_{r=r_{uco}}>0(V''_{eff}< 0 )$.  Here the prime denotes differentiation
with respect to $r$ and $r=r_{uco}$ denotes  the radius for an  unstable photon orbit.
These conditions imply
\begin{eqnarray}
[(r^2_{uco}+a^2)-a\xi]^2- \triangle(r_{uco})[\eta + (\xi-a)^2] &=& 0\,,  \label{fiscon} \\
4r_{uco}[(r^2_{uco}+a^2)-a\xi]-\triangle'(r_{uco})[\eta + (\xi-a)^2] &=& 0 \,. \label{twocon}
\end{eqnarray}
After eliminating $\eta$ from (\ref{fiscon}) and (\ref{twocon}) and solving for $\xi$, we obtain
\begin{eqnarray}
\xi=\frac{r^2_{uco}+a^2}{a},~~~ \xi= \frac{(r^2_{uco}+a^2)\triangle'(r_{uco})-4r_{uco}
\triangle(r_{uco})}{a\triangle'(r_{uco})} \label{weobt}  \,.
\end{eqnarray}
One of these is  necessary  for  describing a black hole shadow.
The first is not suitable for  describing the black hole shadow because it represents principal null congruences.
Taking the second solution, we solve for $\eta$  from Eq.\ (\ref{twocon}) to give
\begin{eqnarray}
\eta&=& \frac{r^2_{uco}[16a^2 \triangle_{uco} -(r_{uco}\triangle'_{uco}-4\triangle_{uco})^2 ]}{a^2 \triangle'^2_{uco} } \,. \label{etacon}
\end{eqnarray}
We note that $\xi$ and $\eta$ of unstable photon orbits describe the contour of a shadow.
Explicitly, the unstable photon orbits are related to the boundary of a shadow.
The apparent shape of a shadow is  obtained by making use of the
 coordinates $\alpha$ and $\beta$  lying in the celestial plane perpendicular to the line joining the observer and the center of spacetime geometry.  The coordinates
$\alpha$ and $\beta$ are found to be~\cite{Bardeen:1973tla,Vazquez:2003zm,Cunha:2016bpi}
\begin{eqnarray}
\alpha &=& \lim_{r_o \rightarrow \infty} \Big[-r^2_o \sin\theta_o \frac{d\phi}{d r}|_{(r_o, \theta_o)} \Big]
 = - \frac{\xi}{\sin\theta_o}  \,, \nonumber \\
\beta &=& \lim_{r_o \rightarrow \infty} \Big[r^2_o \frac{d\theta}{d r}|_{(r_o, \theta_o)} \Big]=
\pm \sqrt{\eta + a^2 \cos^2 \theta_o - \xi^2 \cot^2\theta_o}\,,
\label{shapara}
\end{eqnarray}
where $(r_o, \theta_o)$ denote  the observer's position. A line joining the origin with the observer
is normal to the $\alpha\beta$-plane. Approximately, it is an angular radius of shadow in two orthonormal directions.
From Eq.\ (\ref{shapara}), one can obtain
\begin{equation}
(\alpha - a\sin\theta_o)^2 + \beta^2 = (a+\xi)^2+\eta \,, \label{paraeq}
\end{equation}
which  represents the rim of
the black hole reconstructed from the light rays in the unstable orbit.  If $\theta_o=0$, the shadow appears spherical.

For simplicity,  requiring that the observer be  located in the equatorial plane
$(\theta_o=\pi/2)$, $\alpha$ and $\beta$ are directly related to $\xi$ and $\eta$ as
\begin{eqnarray}
\alpha =-\xi,~~~ \beta= \pm \sqrt{\eta}\,.
\end{eqnarray}

\subsection{Shadow observables \label{sec4-2-1}}

We  may use  four shadow observables to extract  the information on parameters of a black hole~\cite{Takahashi:2004xh, Hioki:2009na, Abdujabbarov:2016hnw, Tsupko:2017rdo, Kumar:2018ple}.
These include  the approximate shadow radius $R_s$, distortion parameter $\delta_s$,
area of the shadow $A_s$, and oblateness $D_{os}$. We  could
determine the  mass,  rotation (spin), and  inclination angle. Four observables take the forms
\begin{eqnarray}
R_s =\frac{(\alpha_r-\alpha_t)^2+\beta^2_t}{2|\alpha_r-\alpha_t|}\,,~~~ \delta_s = \frac{D_{cs}}{R_s} =\frac{ |\tilde{\alpha}_l - \alpha_l| }{R_s}\,,~~~ A_{s} =2\int^{r_{max}}_{r_{min}} \beta \alpha' dr \,,~~~ D_{os} =\frac{\alpha_r - \alpha_l}{\beta_t - \beta_b}\,.
\label{observables}
\end{eqnarray}
Making using of {\it Mathematica} program based on Eqs.\ (\ref{shapara}) and (\ref{observables}), we obtain the following figures.

\begin{figure}
\begin{center}
\subfigure[Schematic illustration for Eq.\ (\ref{observables}) ]
{\includegraphics[width =3. in]{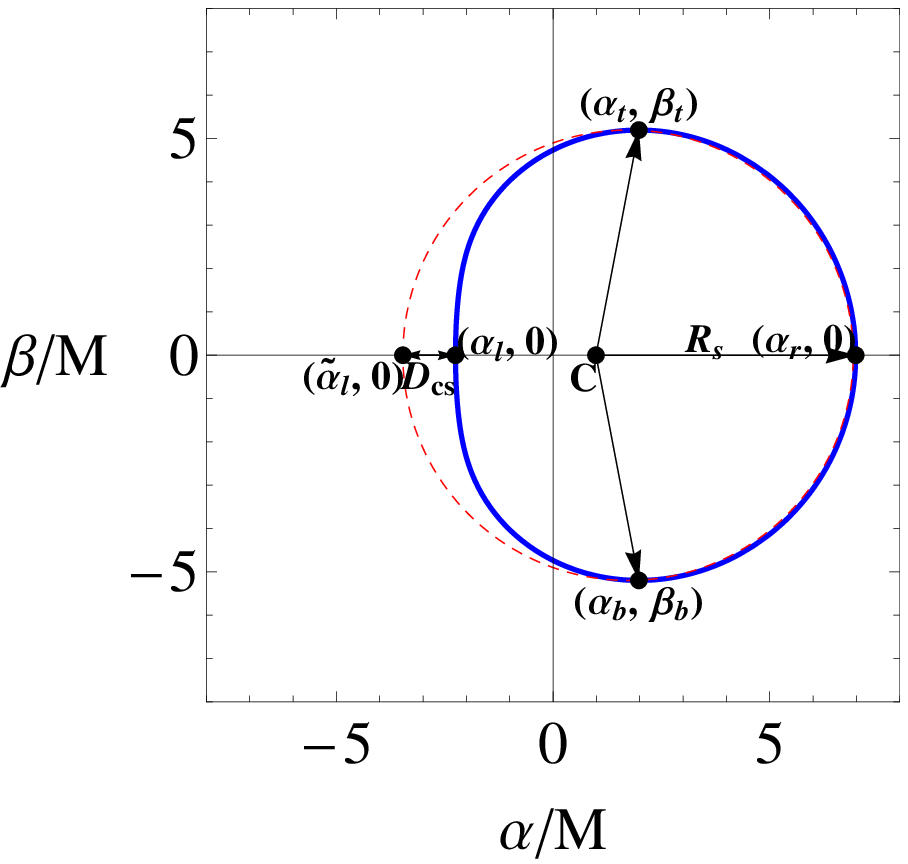}}
\subfigure[Kerr one with increasing $a$]
{\includegraphics[width =3. in]{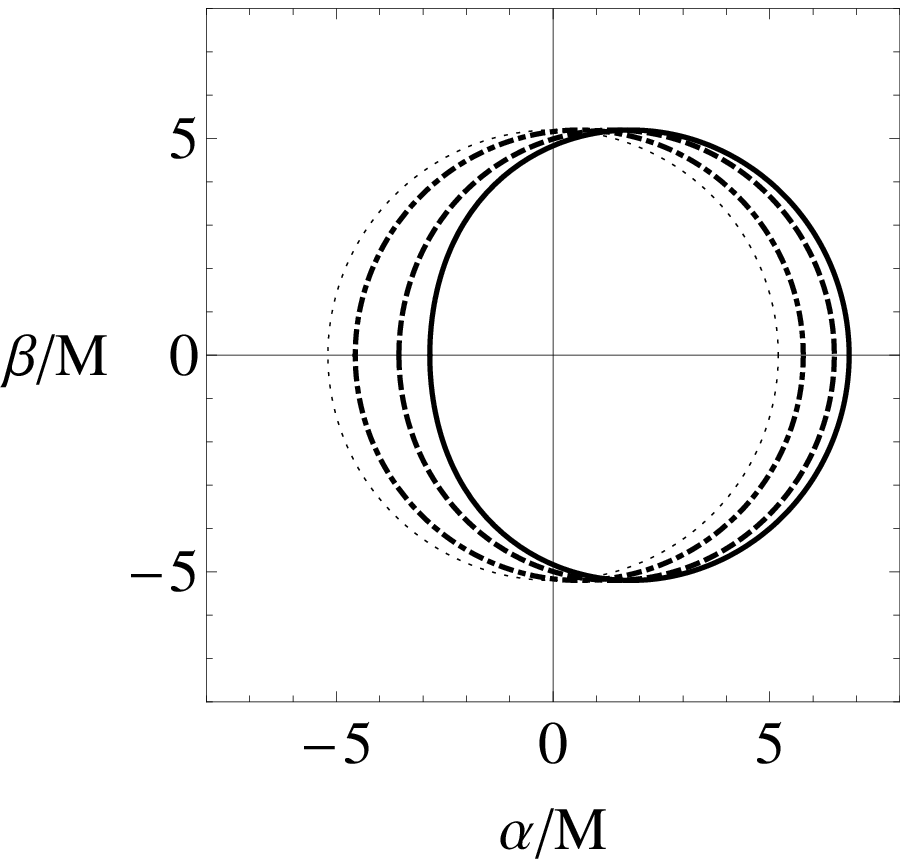}}\\
\subfigure[Kerr one with increasing $\theta$]
{\includegraphics[width =3. in]{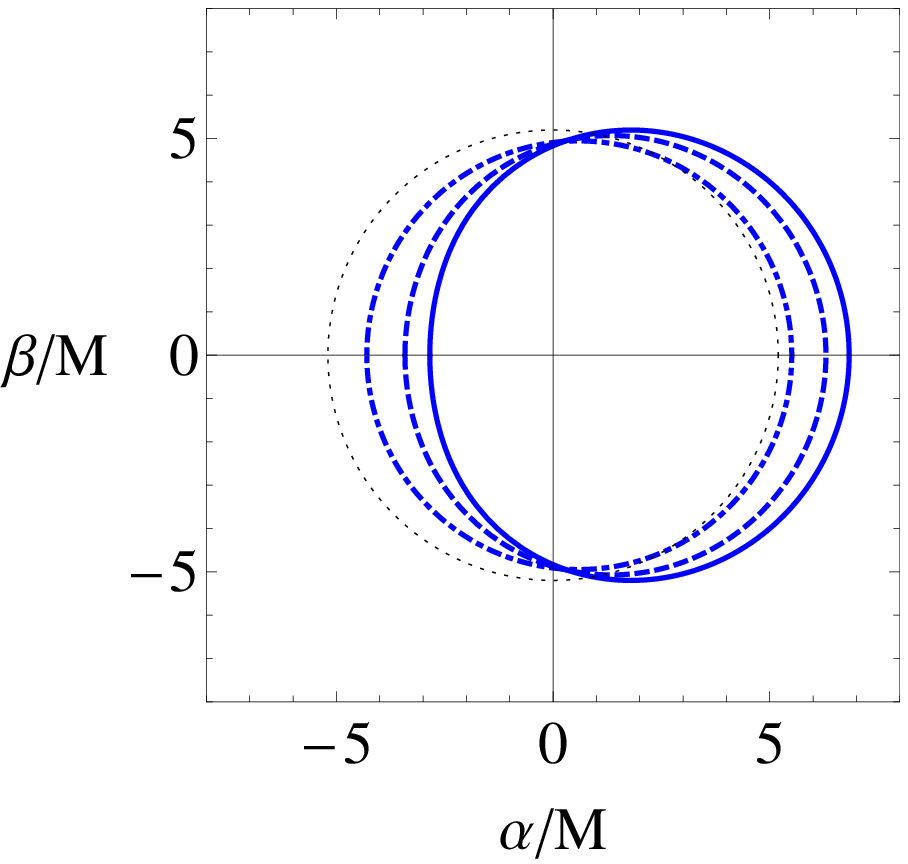}}
\subfigure[Kerr-Newman one with increasing $Q$]
{\includegraphics[width =3. in]{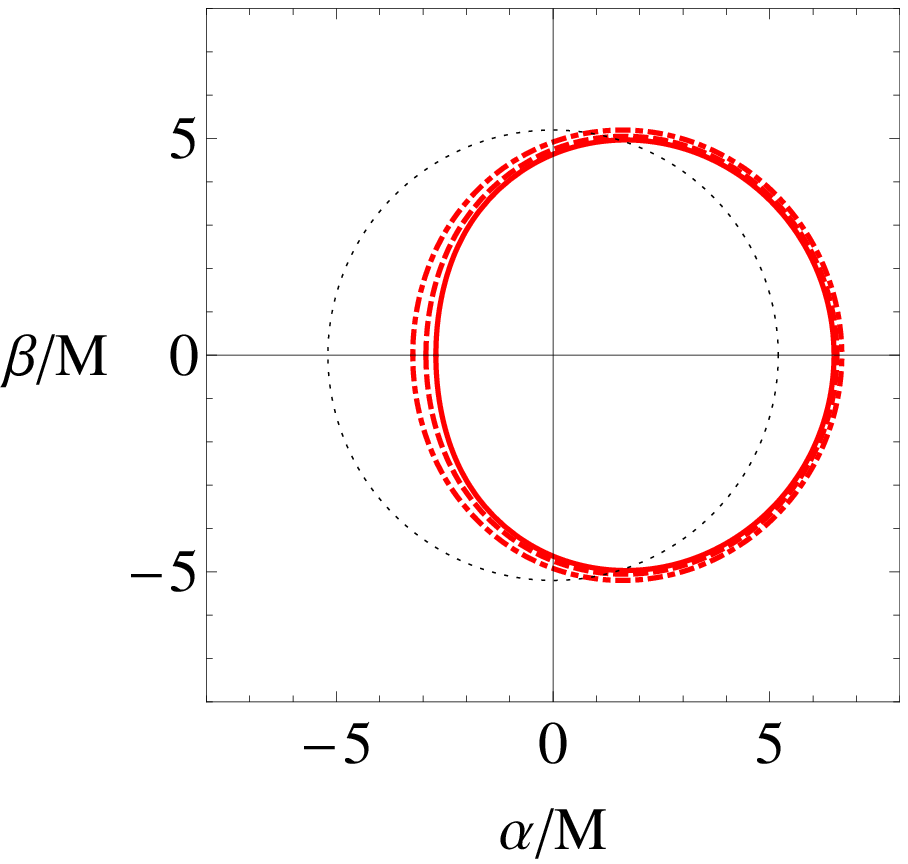}}
\end{center}
\caption{\footnotesize{(color online). Schematic illustration for  observables of a black hole shadow. }}
\label{Schema-illu}
\end{figure}
Figure \ref{Schema-illu} represents schematic illustrations of a black hole shadow.
The rotation direction  with $a > 0$ is assumed to be  counterclockwise   when observing from the positive $\beta$-axis. The closed asymmetric circles represent
the gravitational capture cross section of photons. The asymmetry occurs  because the locations of
unstable orbits for corotating and counterrotating cases are different [See Fig.\ \ref{Effective_potential}(a)].
As we mentioned in Sec.\ $(3)$, the number of photons passing through the left side of the black hole
rotation axis is different from that through  right side. Consequently, it is found that the left side of the black hole is brighter than the right side.

In Fig.~\ref{Schema-illu}(a), $D_{cs}$ denotes  the difference between
the left endpoints of the shadow and the reference circle. The subscript $t$, $r$, $l$ indicate, respectively,
the coordinates of the shadow vertices at the top, right, and left endpoint, while ${\tilde \alpha}_l$ is the coordinate
for the left edge of  referenced circle.  $R_s$ is the approximate shadow radius of the circle passing through
three points, $\alpha_t$, $\alpha_b$, and $\alpha_r$.  $(\alpha_r,\alpha_l)$ and $(\beta_t,\beta_b)$ represent the
horizontal and vertical diameters of the shadow, respectively.

Figure \ref{Schema-illu}(b) shows the shadow of Kerr black hole
with increasing $a$. The grey dotted circle indicates the shadow induced by Schwarzschild black hole  with $a=0$. The black dot-dashed
circle shows the shadow induced by Kerr black hole  with $a=0.3$, the dashed circle with $a=0.7$, the solid circle with $a=0.9$,
respectively. The circle's vertical-axis asymmetry increases and shifts to the right as  $a$ increases.

Figure \ref{Schema-illu}(c) implies  the shadow of Kerr black hole with increasing $\theta$ with $a=0.9$.
The blue dot-dashed circle shows the shadow by Kerr black hole  with $\theta=17.14$, the dashed circle with $\theta=\pi/4$,
the solid circle with $\theta=\pi/2$, respectively. The circle's vertical-axis asymmetry increases and shifts to the right,
as $\theta$ increases.

Figure \ref{Schema-illu}(d) indicates  the shadow of Kerr-Newman black hole with
increasing $Q$. The red dot-dashed circle shows the shadow induced  by Kerr-Newman black hole  with $a=0.8$ and $Q=0$ (Kerr black hole),
the dashed circle with $Q=0.4$, the solid circle with $Q=0.5$, respectively.
As $Q$ increases, the circle's vertical-axis asymmetry increases and shifts to the right, and the area decreases.

From now on, we consider the case that the observer is located in the equatorial plane $(\theta_o=\pi/2)$.

\begin{figure}[H]
\begin{center}
\subfigure[$w$=3/4 and $Q$=0]
{\includegraphics[width =3. in]{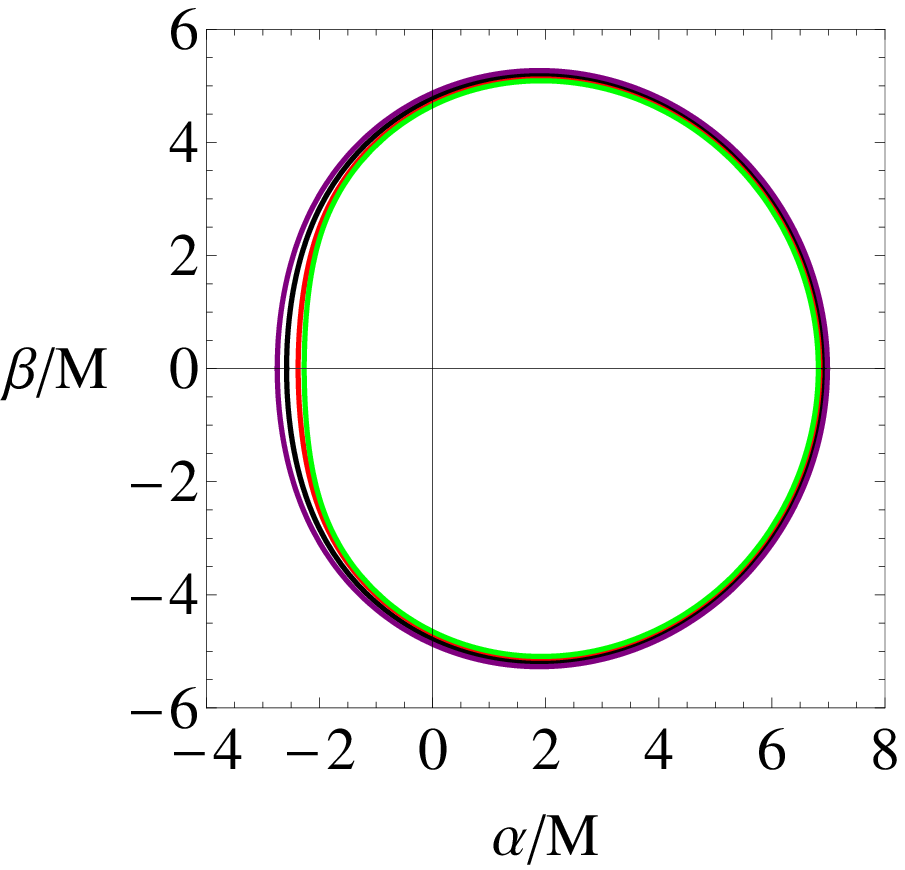}}
\subfigure[$w$=6/4 and $Q$=0]
{\includegraphics[width =3. in]{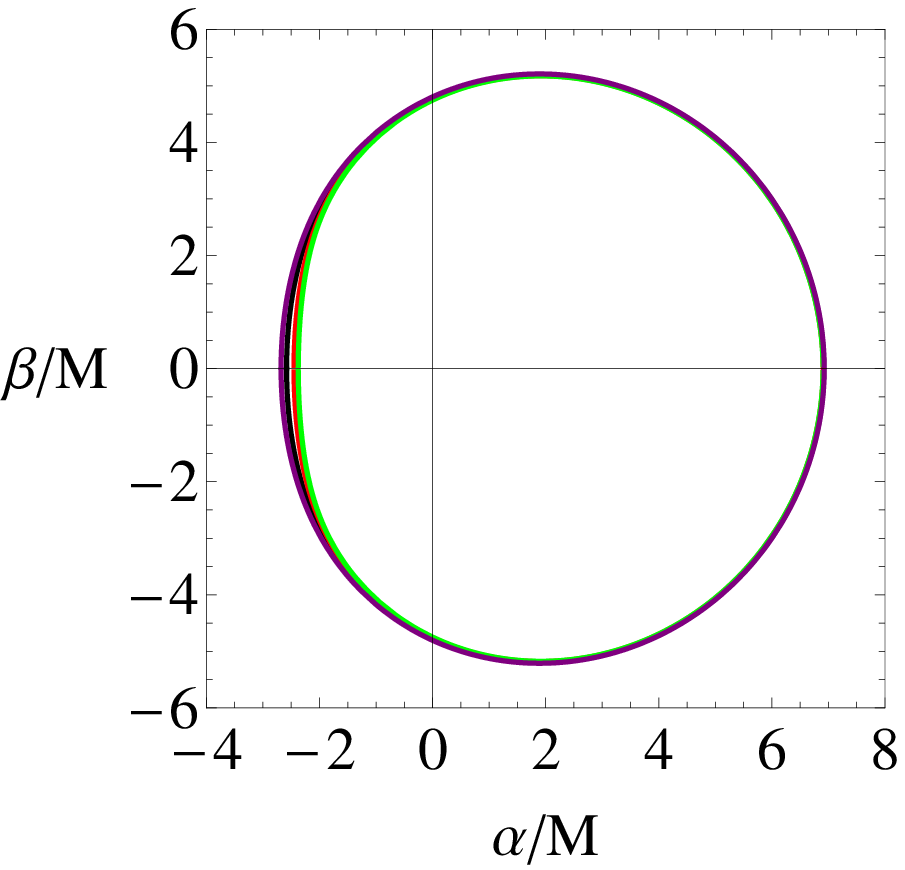}}\\
\subfigure[$w$=3/4 and $Q$=0.15]
{\includegraphics[width =3. in]{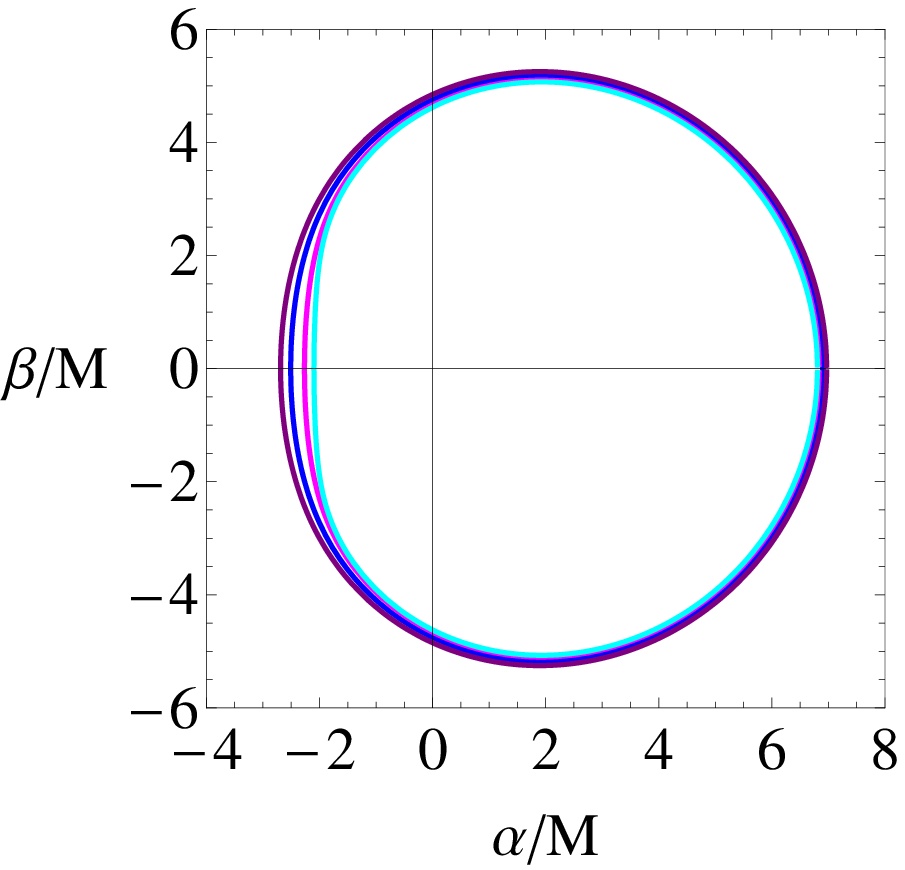}}
\subfigure[$w$=6/4 and $Q$=0.15]
{\includegraphics[width =3. in]{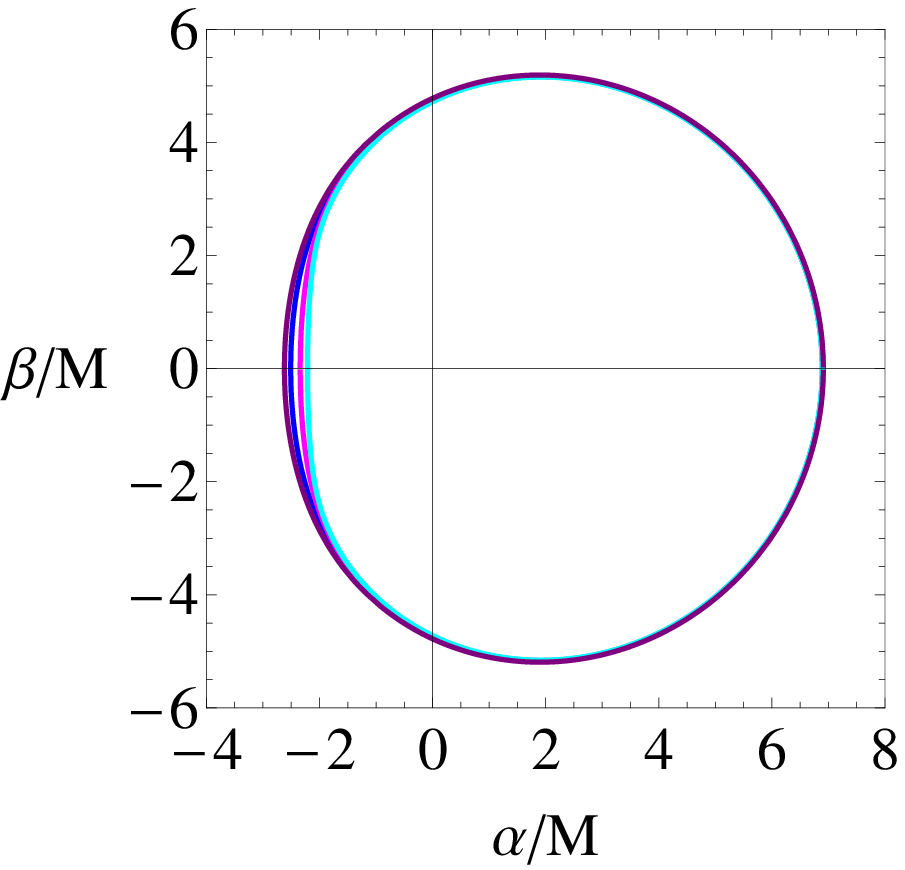}}
\end{center}
\caption{\footnotesize{(color online). Shadow casts induced  by rotating black holes with anisotropic matter. (a) and (b) Kerr black hole  with $Q=0$. (c) and (d) Kerr-Newman black hole with $Q=0.15$. }}
\label{shadow cast}
\end{figure}
Figure \ref{shadow cast} represents the shadow casts by rotating black holes with  selected  parameters.
Figure \ref{shadow cast}(a) and (b) are shadow casts with $Q=0$, $w=3/4$ for (a) and $w=6/4$ for (b), while (c) and (d) are those
with $Q=0.15$, $w=3/4$ for (c) and $w=6/4$ for (d), respectively.
In Fig.\ \ref{shadow cast}(a) and (b), the black circles show shadow casts by Kerr black hole, the red circles show those with $K=-0.05$,
the green circles show those with $K=-0.07$, and the purple circle shows those with $K=0.05$.
In Fig.\ \ref{shadow cast}(c) and (d), the blue circles show shadow casts induced by Kerr-Newman black hole,
the magenta circles show those with $K=-0.05$, the cyan circles show those with $K=-0.07$, and the purple circle shows those
with $K=0.05$. The circle's vertical-axis asymmetry increases and it shifts to the right, and the area increases as $K$ increases.
The difference among the deformed circles decreases as $w$ increases. Because the positive energy condition,
$Q^2+r^{2w}_o r^{2(1-w)}\geq 0$ as shown in \cite{Kim:2019hfp}, for the cases with $K=0.05$, we focus on figuring out
the shadow cast depending on the negative $K$.

\begin{figure}[H]
\begin{center}
\subfigure[Shadow radius $Q$=0]
{\includegraphics[width=3. in]{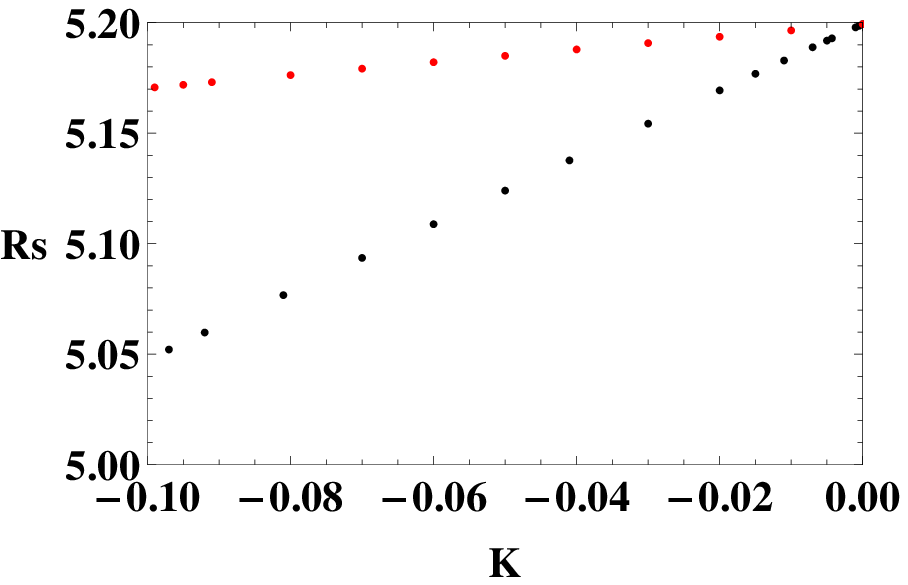}}
\subfigure[Shadow radius $Q$=0.15]
{\includegraphics[width=3. in]{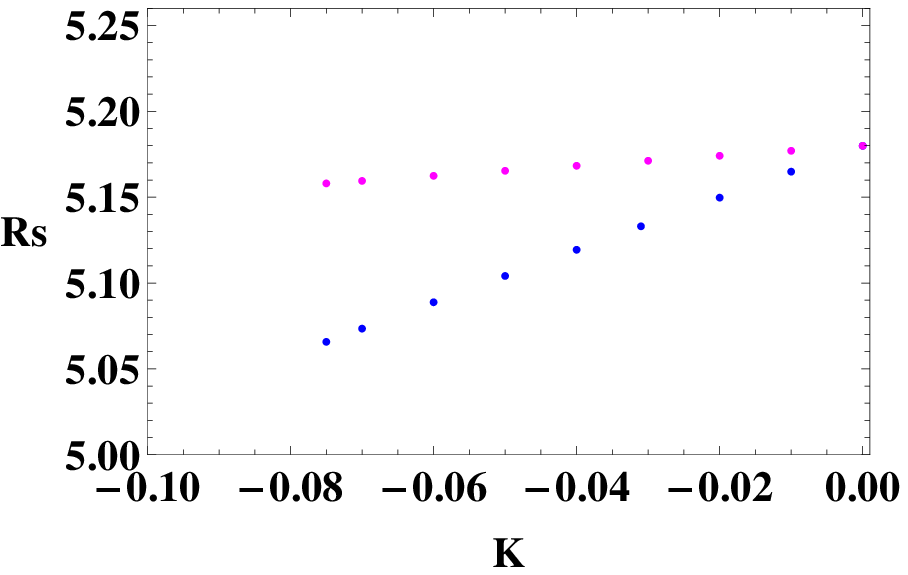}}
\end{center}
\caption{\footnotesize{(color online).
Approximate shadow radius $R_s$ as function of $K$.}}
\label{app_sha_rad}
\end{figure}

Figure \ref{app_sha_rad} represents the approximate shadow radius $R_s$  with respect to $K$. Figure \ref{app_sha_rad}(a) is the case  with
$a=0.95M$ and $Q=0$. The black dots denote  the radius with $w=3/4$, while the red dots represent  the
radius with $w=6/4$.  Figure \ref{app_sha_rad}(b) is the case  with $a=0.95M$ and $Q=0.15M$.
The blue dots correspond to the radius with $w=3/4$, while the magenta dots correspond
to the radius with $w=6/4$. The graphs show that the approximate shadow radius
$R_s$ increases as $K$ increases.

\begin{figure}[H]
\begin{center}
\subfigure[Distortion parameter $Q$=0]
{\includegraphics[width=3. in]{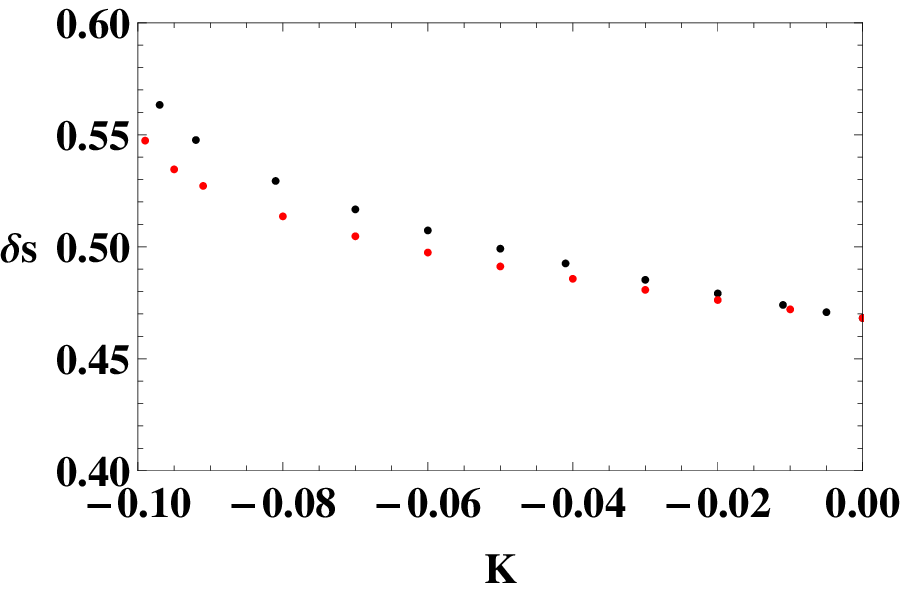}}
\subfigure[Distortion parameter $Q$=0.15]
{\includegraphics[width=3. in]{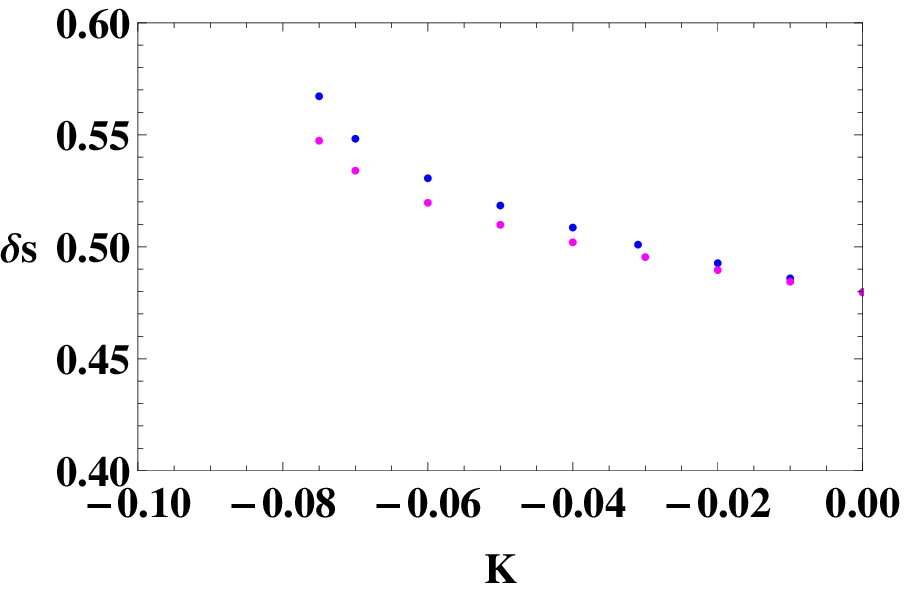}}
\end{center}
\caption{\footnotesize{(color online).
Distortion parameter $\delta_s$ as function of $K$.}}
\label{disp_para}
\end{figure}

Figure \ref{disp_para} represents the distortion parameter $\delta_s$  versus  $K$. Figure \ref{disp_para}(a) is the case  with
$a=0.95M$ and $Q=0$. The black dots correspond to the distortion parameter with $w=3/4$, while the red dots
correspond to the distortion parameter with $w=6/4$. Figure \ref{disp_para}(b) is the case  with $a=0.95M$ and $Q=0.15M$.
The blue dots represent  the distortion parameter with $w=3/4$, while the magenta dots denote  the distortion parameter with $w=6/4$.
These  show that the distortion parameter $\delta_s$ decreases as $K$ increases.

\begin{figure}[H]
\begin{center}
\subfigure[Area $Q$=0]
{\includegraphics[width=3. in]{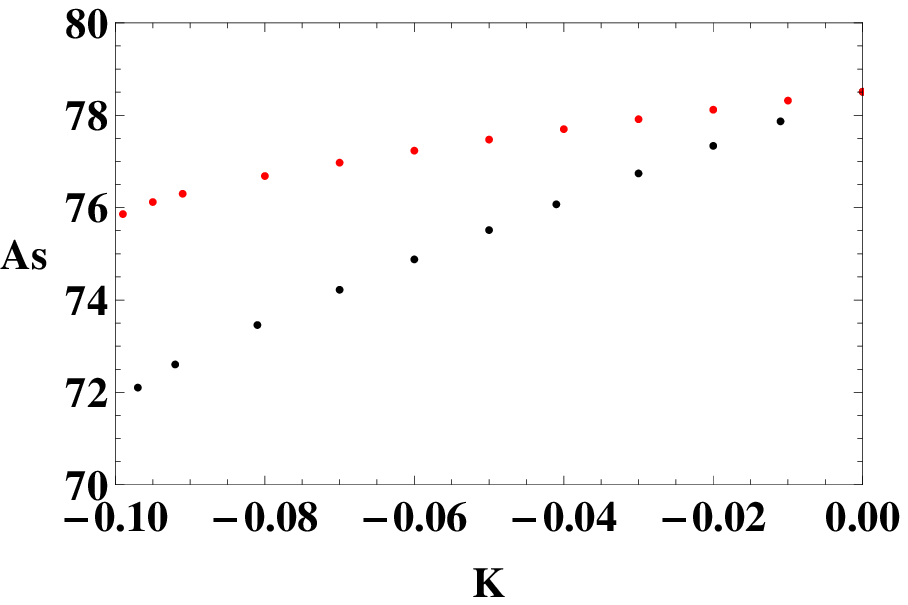}}
\subfigure[Area $Q$=0.15]
{\includegraphics[width=3. in]{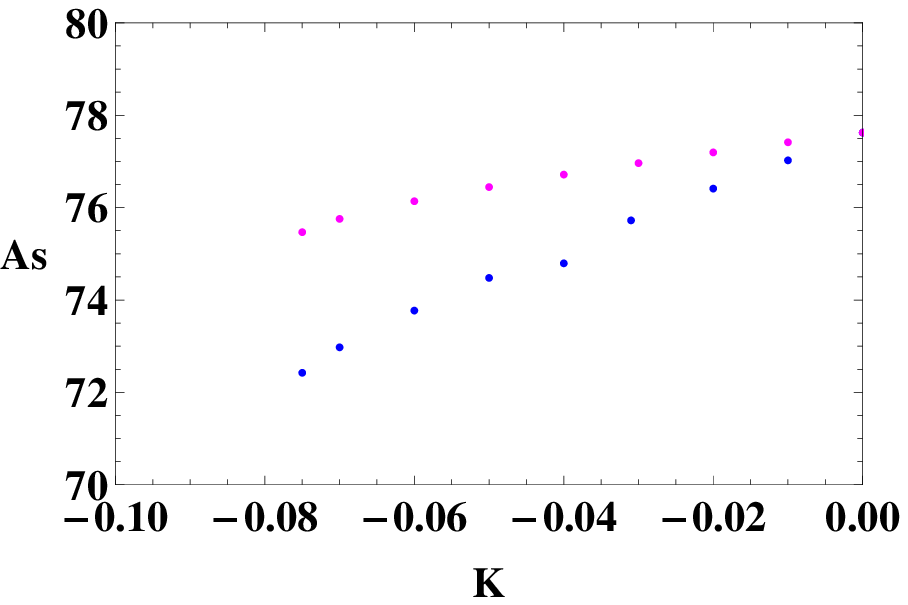}}
\end{center}
\caption{\footnotesize{(color online).
Area $A_s$ as function of $K$.}}
\label{sarea}
\end{figure}
Figure \ref{sarea} presents the area  $A_S$ with respect to $K$. Figure \ref{sarea}(a) is the case  with
$a=0.95M$ and $Q=0$. The black dots correspond to the area with $w=3/4$, while the red dots
denote  the area with $w=6/4$. Figure \ref{sarea}(b) is the case  with $a=0.95M$ and $Q=0.15M$.
The blue dots correspond to the area with $w=3/4$, while the magenta dots indicate  the area with $w=6/4$.
The graphs show that the area of the shadow $A_s$ increases as $K$ increases.

\begin{figure}[H]
\begin{center}
\subfigure[Oblateness $Q$=0]
{\includegraphics[width=3. in]{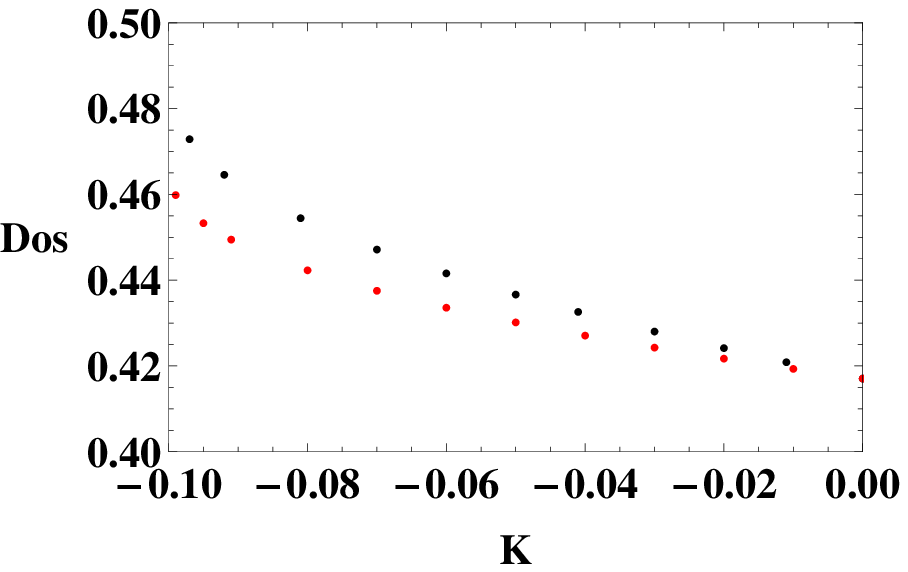}}
\subfigure[Oblateness $Q$=0.15]
{\includegraphics[width=3. in]{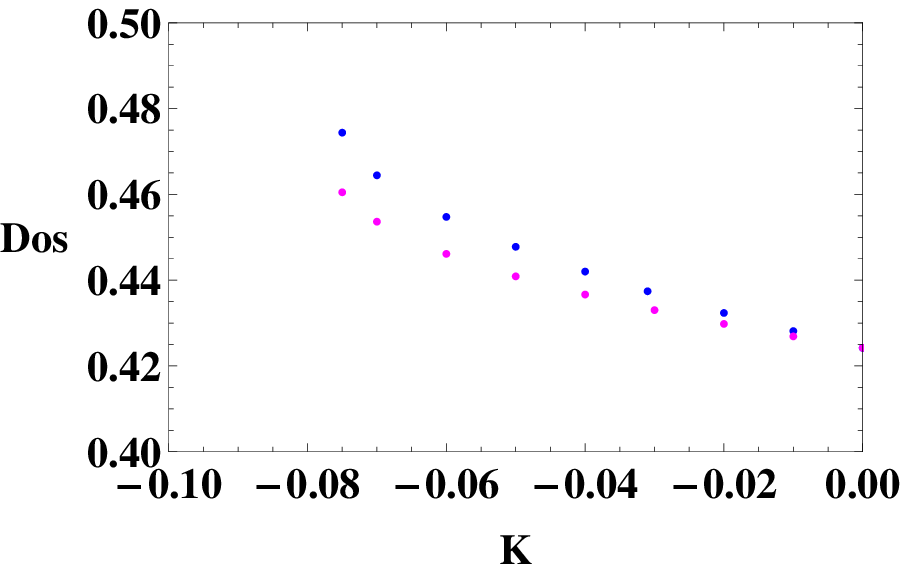}}
\end{center}
\caption{\footnotesize{(color online).
Oblateness $D_{os}$ as function of $K$.}}
\label{oblate}
\end{figure}
Figure \ref{oblate} indicates  the oblateness with respect to $K$. Figure \ref{oblate}(a) is the case  with
$a=0.95M$ and $Q=0$. The black dots correspond to the oblateness  with $w=3/4$, while the red dots
correspond to the oblateness  with $w=6/4$. Figure \ref{oblate}(b) is the case  with $a=0.95M$ and $Q=0.15M$.
The blue dots correspond to the oblateness with $w=3/4$, while the magenta dots correspond to the oblateness with $w=6/4$.
The graphs show that the oblateness $D_{os}$ decreases when $K$ increases.

\section{Summary and discussions \label{sec5}}

We have investigated the shadow cast induced  by a rotating black hole with
an anisotropic matter field. For this study, we assumed that  the matter field interacts with
light rays through  gravitational interaction.
We first explored  the symmetry of the rotating black hole geometry  and its  separability structure.
We have found the Killing tensor explicitly, implying
the separability structure and  integrability of the geodesic motions.

The geodesic equations are derived by adopting the Hamilton-Jacobi formalism.
We have analyzed the radial null geodesic motion for both corotating and counterrotating cases.
The photon orbits for corotating cases are located inside the ergosphere, implying
that some counterrotating photons can change their rotating direction.
Thus, it is found that one side of the black hole  is  brighter than the other side.
We used  the backward ray-tracing algorithm to get the relation
between two impact parameters and the coordinates axes in spatial infinity.
The size of the black hole shadow depends on its mass mainly, while the shape  depends   both rotation
and inclination angle. In this work, we have taken  $\theta=\pi/2$, large values of $Q$ and $K$
to show  the difference among  parameters clearly. Also, we selected specific values $w$.
We expect that densities for $Q$ and $K$ are very smaller
than those for  mass and  rotation for a real black hole.

We have presented  the shadow cast induced by rotating black holes in Fig.\ \ref{shadow cast}. The left side corresponds to
the corotating direction between the light rays and the black hole,
while the right side to the counterrotating one between the light rays and the black hole.
It is observed that the left side becomes flattened when $K$ decreases.
We have investigated four observables to see their dependence on parameters of a black hole.
The approximate shadow radius $R_s$ increases as  $K$ increases.
The distortion parameter $\delta_s$ decreases as $K$ increases
and $w$ has the lower value for both cases.
The area of the shadow $A_s$ increases as $K$ increases.
Lastly, the oblateness $D_{os}$ decreases as  $K$ increases.
If we require the positive energy
condition satisfying $Q^2+r^{2w}_o r^{2(1-w)}\geq 0$ as shown in \cite{Kim:2019hfp},
there will be a small room for the positive value of $K$. Therefore, we mainly analyzed four observables
with negative values of $K$, except for Fig.\ \ref{shadow cast} to get the deformation behavior of
the shadow cast. Our results suggest that the observed black hole mass
might be either underestimated or overestimated depending on the sign of the matter field ($K$),
even though the difference is  very small.

According to~\cite{Akiyama:2019cqa}, they measured  emission ring diameter   $d=42\pm3 \mu$as,
angular size  $11^{+0.5}_{-0.3}$ in units of mass, and  observed inclination
angle $\theta_{o}\approx 17^\circ$~\cite{Walker:2018muw, Akiyama:2019cqa} for  M$87^*$'s shadow.
It suggests that the supermassive black hole (M$87^*$)
is lying at a small angle in the direction of the
observer's line of sight. Thus, the angular size seems to be similar to that for Schwarzschild black hole.
However, one knows that  M$87^*$ rotates with an estimated rotation $a=0.90\pm0.05$ \cite{Tamburini:2019vrf}.
It is worth to note   that a significant difference between  position angles of brightness maximum
measured in $2013$ and $2017$ was found in~\cite{Wielgus:2020zvj}.
On the other hand, there have been attempted to make an  image of  SgrA$^*$ in the  Milky
Way~\cite{Balaps1974, Ghez:2008ms, Gillessen:2008qv}
in the radio spectrum through very-long-baseline interferometry (VLBI)
experiments \cite{Falcke:1999pj, Goddi:2017pfy}. They estimated the  emission ring diameter  for the source as
$d\sim 50 \mu$as\cite{Lu:2018uiv,Ghez:2008ms, Gillessen:2008qv}
and the inclination angle $\theta_o >30^\circ$~\cite{Broderick:2008sp,Reid:2014boa}.
It is interesting to note that SgrA$^*$ has a larger inclination angle than M87$^*$.
Hence,  one expects that  its shadow cast will show an asymmetric brighter side  than  M$87^*$ and be detected with new-generation instruments.

Two research directions for the black hole shadow
are known as theoretical and observational approaches.
In this paper, we have focused on the theoretical aspect in light of the observation.
At this stage, it is not easy  to determine the  parameters of the rotating black hole exactly  through
the shadow image. However, we hope that both the upgraded EHT and the BlackHoleCam project will
detect the shadow image  with a much higher resolution in the near future.

Finally, we did not consider the effect of an accretion disk at all. For this,
we should  construct the accretion disk \cite{Novikov:1973kta, Page:1974he, Jang:2020abp, Collodel:2021gxu}
in the geometry of the rotating black hole with an
anisotropic matter field. This issue will be considered as an interesting subject and thus,  we remain it for a future work.

\section*{Acknowledgments}

B.-H.\ L. (NRF-2020R1F1A1075472), W.\ L. (NRF-2016R1D1A1B01010234), Y.\ S.\ M. (NRF -2017R1A2B4002057), and
Center for Quantum Spacetime (CQUeST) of Sogang University (NRF -2020R1A6A1A03047877) were supported by Basic
Science Research Program through the National Research Foundation of Korea funded by the Ministry of Education.
We would like to thank Gungwon Kang, Inyong Cho, Wontae Kim, and Myeong-Gu Park for helpful discussions and comments, and
thank Hyeong-Chan Kim and Youngone Lee for their hospitality during our visit to Korea National University of
Transportation, and Jin Young Kim to Kunsan National University.
We would like to thank Yoonbai Kim and O-Kab Kwon for their hospitality at SGC 2020 held in Pohang.
WL also would like to thank Seyen Kouwn and Hocheol Lee for helping the numerical work.

\end{document}